\documentclass[aps,superscriptaddress,prl,amsmath,preprint]{revtex4}
\usepackage{amssymb}
\usepackage{graphicx,subfigure,float}% Include figure files
\usepackage[percent]{overpic}
\usepackage{dcolumn}% Align table columns on decimal point
\usepackage{bm}% bold math
\usepackage{color}
\usepackage{hyperref}% add hypertext capabilities
\begin{document}
%++++++++++++++++++++++++++++++++++++++++++++++++++++
\preprint{APS/123-QED}
%++++++++++++++++++++++++++++++++++++++++++++++++++++
%++++++++++++++++++++++++++++++++++++++++++++++++++++
\title{Intrinsic Spin Hall Conductivity Platform in Triply Degenerate Semimetal}
%============================================
\author{Zhengchun Zou$^\ddagger$}
 \affiliation{Hunan Provincial Key laboratory of Thin Film Materials
and Devices, Xiangtan University, Xiangtan 411105, China}
 \author{Pan Zhou$^\ddagger$}
 \email{zhoupan71234@xtu.edu.cn}
\affiliation{School of Material Sciences and Engineering, Xiangtan
University, Xiangtan 411105, China}
\author{Rui Tan}
 \affiliation{Hunan Provincial Key laboratory of Thin Film Materials
and Devices, Xiangtan University, Xiangtan 411105, China}
\author{Wenqi Li}
 \affiliation{Hunan Provincial Key laboratory of Thin Film Materials
and Devices, Xiangtan University, Xiangtan 411105, China}
 \author{Zengsheng Ma}
\affiliation{School of Material Sciences and Engineering, Xiangtan
University, Xiangtan 411105, China}
\author{Lizhong Sun}
 \email{lzsun@xtu.edu.cn}
 \affiliation{Hunan Provincial Key laboratory of Thin Film Materials
and Devices, Xiangtan University, Xiangtan 411105, China}
%============================================
%\keywords{Topological insulator, Bilayer hexagonal lattices, Edge states, Band inversion}
%====================================================================
\date{\today}
%============================================
\begin{abstract}
It is generally believed that conductivity platform can only exist in insulator with topological nontrivial bulk occupied states. Such rule exhibits in two dimensional quantum (anomalous) Hall effect, quantum spin Hall effect, and three dimensional topological insulator. In this letter, we propose a spin Hall conductivity (SHC) platform in a kind of three dimensional metallic materials with triply degenerate points around the Fermi level. With the help of a four bands \textbf{k}${\cdot}$\textbf{p} model, we prove that SHC platform can form between $|\frac{3}{2},\pm\frac{3}{2}\rangle$ and $|\frac{1}{2},\pm\frac{1}{2}\rangle$ states of metallic system. Our further ab initio calculations predict that a nearly ideal SHC platform exhibits in an experimentally synthesized TaN. The width of the SHC platform reaches up to 0.56 eV, hoping to work under high temperature.  The electrical conductivity tensor of TaN indicates that its spin Hall angle reaches -0.62, which is larger than many previous reported materials and make it an excellent candidate for producing stable spin current.\\
\end{abstract}
%====================================================================
%\pacs{71.20.-b, 71.70.Ej, 73.20.At}
\maketitle
%====================================================================
\indent In recent years, spin Hall effect (SHE) receives much attention due to its potential to produce pure spin current for future high speed spintronics devices\cite{she_rmp}. The transverse spin currents mainly derive from the atomic spin-orbit coupling, and intrinsic or extrinsic mechanisms may contribute the spin Hall conductivity (SHC) of real materials\cite{ahe_rmp}. Among them, the large SHC with intrinsic mechanism attracts much attention recently\cite{4d5d,Pt,W3Ta,TaAs,WTe2}. A well-known materials with intrinsic spin Hall effect is 2D topological insulator, in which the SHC can be quantized when the Fermi level is tuned into the band gap and the width of the conductivity platform matches with the value of the gap\cite{qshe_1,qshe_2,qshe_3,qshe_4}. The similar spin Hall platform also appears in 3D topological insulator\cite{Bi2Se3_1} but the conductivity is not necessary to be quantized. For device application, however, the low carrier density in topological insulator make it improper to produce large spin current in concrete electronic devices. We wonder is there any metallic material can produce stable SHC platform independent of the doping concentration.\\
%====================================================================
%+++++++++++++++++++++++++++++++++++++++++++++++++++++++++++++++++++++++
 \begin{figure}
% Requires \usepackage{graphicx}
\includegraphics[trim={0.0in 0.0in 0.0in 0.0in},clip,width=6.0in]{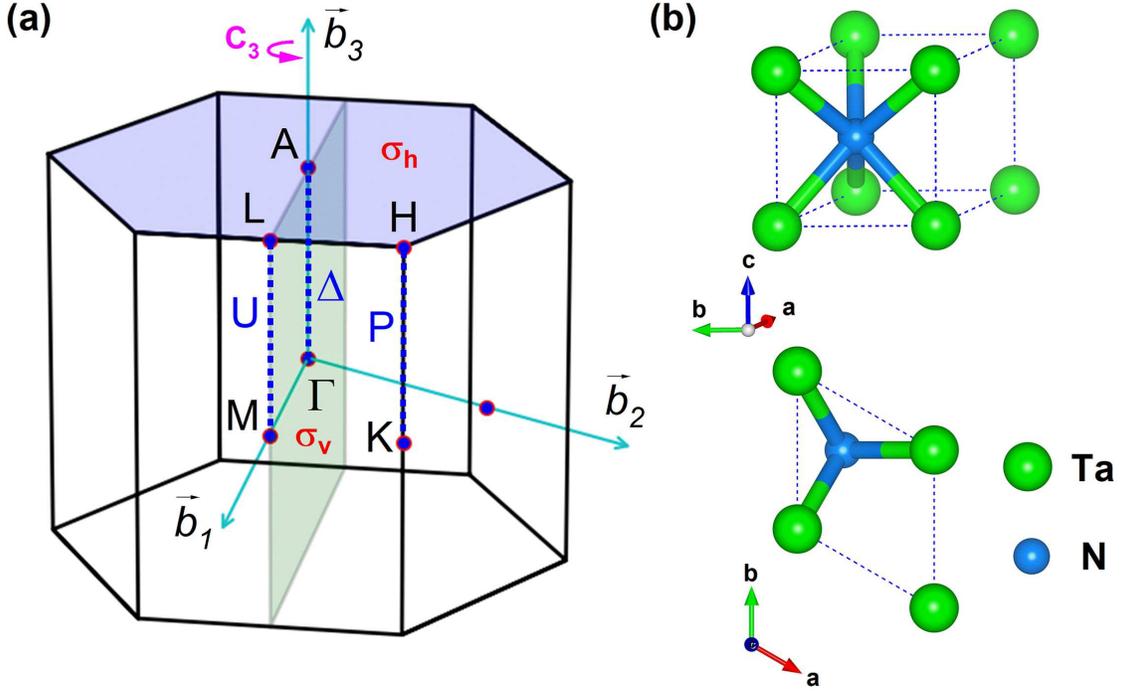}\\
\caption{(color online) (a) is the first Brilloune zone for crystal with space group of P$\bar{6}$m2. High symmetry points and lines are label with the corresponding letter of P$\bar{6}$m2. The C$_3$ rotation axis, $\sigma_h$, and one of the $\sigma_v$ mirror are also presented in the figure. Side (b) and top view (c) of crystal structure of TaN.\\ }\label{fig1}
\end{figure}
%+++++++++++++++++++++++++++++++++++++++++++++++++++++++++++++++++++++++
\indent Finding materials with large SHC is also an important topic in the area of spin Hall effect. Latest works proposed some material families with large SHC, such as A$_{15}$ superconductors \cite{W3Ta,beta-W} and Weyl semimetal\cite{TaAs}. However, the SHC of the materials strongly depends on the concentration of doping and their SHC value changes rapidly with the change of the Fermi level (for example the change produced by the thermal fluctuations in real device). It is a big barrier for the SHE to be applied in future device to produce stable spin current under high temperature. \\
%+++++++++++++++++++++++++++++++++++++++++++++++++++++++++++++++++++++++
\indent In this letter, we propose that a SHC platform can form in triply degenerate semimetals\cite{TaN2016,TaN2016_prx,TDP_science,TDP_exp}. The results obtained from $\textbf{k}\cdot\textbf{p}$ model and first-principles calculations indicate that the SHC platform is mainly contributed by the spin Berry curvature (SBC) between $|\frac{3}{2},\pm\frac{3}{2}\rangle$ and $|\frac{1}{2},\pm\frac{1}{2}\rangle$ states. The SBC distribution around the triply degenerate point (TDP) shows strong localization and characteristics of strong polarized \emph{p} orbitals. Furthermore, the contribution of positive and negative SBC around TDP to the conductivity is severely imbalance forming a platform. To realize the SHC platform in concrete materials, an experimentally synthesized TaN is predicted. The first-principles results confirm that a nearly ideal SHC platform as width as 0.56 eV exhibits in the material and its spin Hall angle reaches up to -0.62.\\
%+++++++++++++++++++++++++++++++++++++++++++++++++++++++++++++++++++++++
 \begin{figure*}
% Requires \usepackage{graphicx}
\includegraphics[trim={0.0in 0.0in 0.0in 0.0in},clip,width=6.4in]{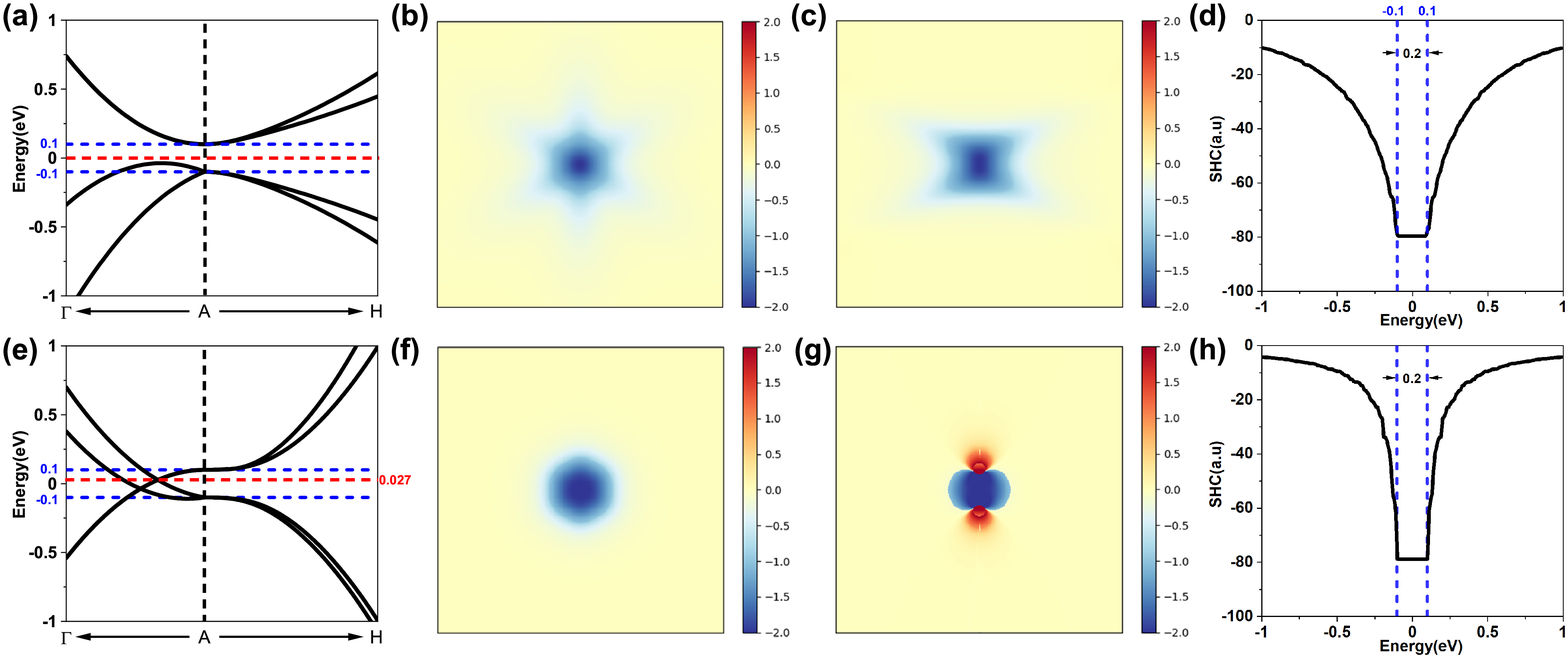}\\
\caption{(color online) Band structures of four-band $k{\cdot}p $ model (without and with triply degenerate point) as well as the SBC and SHC distribution. The momentum zero point is set at A. (a) is the band structure without triply degenerate point. Two blue dashed horizontal lines represent E = $E_F$ - 0.1 eV and E = $E_F$ + 0.1 eV. The k-resolved SBC on a log scale in a slice of the 2D BZ at $k_z$ = 0 (b) and $k_y = 0$ (c) at $E_F$, respectively.  (d) is the SHC $\sigma_{xy}^z$ relative to the Fermi energy. Two blue dashed vertical lines represent E = $E_F$ - 0.1 eV and E = $E_F$ + 0.1 eV, respectively. (e) is the band structure with triply degenerate point. The red dashed horizontal line cross triply degenerate point represent E = $E_F$ + 0.027 eV. The k-resolved SBC on a log scale in a slice of the 2D BZ at $k_z = 0$ (f) and $k_y = 0$ (g) at E = $E_F$ + 0.027 eV , respectively. Where red and blue areas represent positive and negative regions. (h) is the SHC  $\sigma_{xy}^z$ relative to the position of Fermi energy.}\label{fig2}
\end{figure*}
%+++++++++++++++++++++++++++++++++++++++++++++++++++++++++++++++++++++++
\indent The intrinsic SHC can be obtained from $\textbf{k}\cdot\textbf{p}$ model and Wannier Hamiltonian by the Kubo-like formula:
%+++++++++++++++++++++++++++++++++++++++++++++++++++++++++++++++++++++++
\begin{equation}\label{equ1}
\sigma_{ij}^l (\omega)=-\frac{e^2}{\hbar} \sum_{k} \Omega_{ij}^{l}(\textbf{k})
\end{equation}
%+++++++++++++++++++++++++++++++++++++++++++++++++++++++++++++++++++++++
where $\sigma_{ij}^l$ represents the spin current along the $i$ direction generated by an charge current or electric field along the $j$ direction and the spin current is polarized along the $l$ direction. Here $\omega$ is used to consider possible electron or hole doping. The k-resolved SBC can be written as $\Omega_{ij}^{l}(\textbf{k})= \sum_{n} f_{n\textbf{k}} \Omega_{n,ij}^{l}(\textbf{k})$. $f_{n\textbf{k}}$ is the Fermi-Dirac distribution function for the band $n$ at \textbf{k} and the band-projected SBC term is:
%+++++++++++++++++++++++++++++++++++++++++++++++++++++++++++++++++++++++
\begin{equation}\label{equ3}
       \Omega_{n,ij}^{l}(\textbf{k})= \hbar^{2}\sum_{m\neq n}^{}\frac{-2\operatorname{Im}\langle n\textbf{k}|\hat{J}_i^l|m\textbf{k}\rangle
       \langle m\textbf{k}|\hat{v}_j|n\textbf{k}\rangle}{(E_n (\textbf{k})-E_m(\textbf{k}))^2}
\end{equation}
%+++++++++++++++++++++++++++++++++++++++++++++++++++++++++++++++++++++++
 where $\hat{J}_i^l=\frac{1}{2}(\hat{s}_l\hat{v}_i+\hat{v}_i\hat{s}_l)$, $ \hat{v}_j=(1/{\hbar})(\partial\hat{H}/\partial k_j)$.\\
%++++++++++++++++++++++++++++++++++++++++++++++++++++++++++++
\indent In present letter, we concentrate on the SHC in a hexagonal materials with the existence of TDP, which may appear in a C$_3$ rotation invariant axis in reciprocal space and come from the crossing between non-degenerate band and double degenerate bands. The typical space group of this type of material is $P\bar{6}m2 $ (No.187)\cite{TaN2016,Nexus}. At high symmetry point $\Gamma$ or A, the symmetry point group is D$_{3h}$ and the typically electronic states for triply degenerate fermions possibly come from double degenerate $J_z = \pm \frac{1}{2}$ and $\pm \frac{3}{2}$. On high symmetry line $\Delta$ ( $\Gamma$-A), the wave vector point group reduces to C$_{3v}$. The bands with $J_z = \pm \frac{1}{2}$ must be double degenerate restricted by vertical mirror symmetry and $J_z = \pm \frac{3}{2}$ must separate with each other and form two non-degenerate band. When the two groups of bands with the $J_z = \pm \frac{1}{2}$ and $\pm \frac{3}{2}$ cross each other, TDPs may form in C$_3$ rotation invariant axis. In time-reversal symmetry ($T$) system, space inversion symmetry ($P$) must be broken because $PT$ operator force each band double degenerate, which makes TDP impossible to appear in any $\textbf{k}$ points.\\
%++++++++++++++++++++++++++++++++++++++++++++++++++++++++++++
\indent To describe the electronic states around TDP, a four-band minimal $\textbf{k}{\cdot}\textbf{p}$ model was constructed from the point group D$_{3h}$ including C$_3$ rotation operator, one perpendicular mirror ($\sigma_h$), and three vertical mirror operator($\sigma_v$), as shown in Fig. \ref{fig1} (a). Time reversal symmetry must be considered for the time reversal invariant points $\Gamma$ and A. With the basis of $|\frac{3}{2},\pm\frac{3}{2}\rangle$ and $|\frac{1}{2}, \pm\frac{1}{2}\rangle$\cite{group_dis}, we can construct the following $ \textbf{k}{\cdot}\textbf{p}$ Hamiltonian:
%++++++++++++++++++++++++++++++++++++++++++++++++++++++++++++
\begin{equation}
H(\textbf{k}) = \begin{pmatrix}M_1(\textbf{k}) & A_1k_z &  N_{-}(\textbf{k}) & A_4k_{+}k_z \\ A_1k_z & M_1(\textbf{k}) & -A_4k_{-}k_z & N_{+}(\textbf{k})  \\N_{+}(\textbf{k}) & -A_4k_{+}k_z & M_2(\textbf{k}) & 0 \\ A_4k_{-}k_z & N_{-}(\textbf{k}) & 0 & M_2(\textbf{k}) \end{pmatrix}
\end{equation}
%++++++++++++++++++++++++++++++++++++++++++++++++++++++++++++
where $k_{\pm}=k_y \pm ik_x$, $N_{\pm}(\textbf{k})={\pm}A_2k_{\pm} + A_3k_{\mp}^2$, $M_1(\textbf{k}) = \varepsilon_1+B_1k_z^2 +B_2k_{+}k_{-}$,$M_2(\textbf{k}) = \varepsilon_2 + C_1k_z^2 +C_2k_{+}k_{-}$. The detailed process to construct the model can be found in supplemental materials\cite{group_dis}.\\
%++++++++++++++++++++++++++++++++++++++++++++++++++++++++++++++++++++++++
\indent Although following the above symmetry and state conditions, the TDP only possibly appear. Firstly, we consider the two group of bands $|\frac{3}{2},\pm\frac{3}{2}\rangle$ and $|\frac{1}{2}, \pm\frac{1}{2}\rangle$\cite{group_dis} form topological insulator, namely the two group of bands do not cross each other. With proper coefficients in the above Hamiltonian as shown in the supplemental materials\cite{group_dis}, the two groups of bands separate with each other and form global band gap, as shown in Fig. \ref{fig2} (a). To ensure the system topological non-trivial, through adjusting the on-site energies and the quadratic term coefficients, we set the energy of $|\frac{3}{2},\pm\frac{3}{2}\rangle$ lower than that of $|\frac{1}{2}, \pm\frac{1}{2}\rangle$, which is the typical case of semimetals with TDP. The energy band structure, SBC, and SHC are presented in Fig. \ref{fig2} (a)-(d). We find that the SBC mainly localizes around symmetry line $\Gamma$-A, especially around A point as shown in Fig. \ref{fig1} (b) and (c). When Fermi level locate in the valence bands or conduction bands, the SBC drops dramatically due to the severe dispersion of band around the A point.  When the Fermi level is located in the band gap, a electronic conductivity platform forms for SHC as shown in Fig. \ref{fig2} (d) due to the strong localization of the SBC.\\
%++++++++++++++++++++++++++++++++++++++++++++++++++++++++++++++++++++++++
\begin{figure}
% Requires \usepackage{graphicx}
\includegraphics[trim={0.0in 0.0in 0.0in 0.0in},clip,width=5.5in]{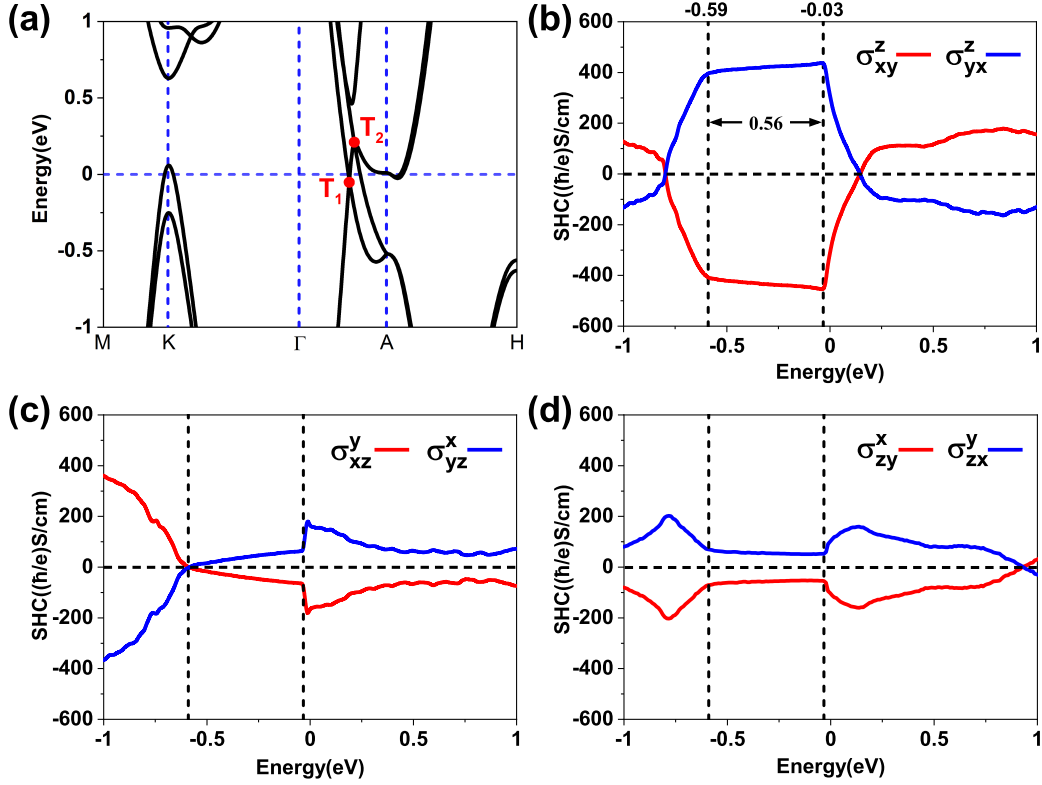}\\
\caption{(color online) (a) Band structure of TaN along high symmetry lines with SOC. Two band crossings T$_{1}$ and T$_{2}$ are signed by red dots. (b)-(d) SHC tensor elements in function of the Fermi energy of TaN. Two black dashed vertical lines represent E = $E_F$ - 0.03 eV and E = $E_F$ - 0.59 eV. }\label{fig3}
\end{figure}
%++++++++++++++++++++++++++++++++++++++++++++++++++++++++++++++++++++++++
\indent Secondly, we investigate the properties when the system is triply degenerate semimetal. According to the Hamiltonian (3), the energy band structure with triply fermion can form on the A-$\Gamma$ high symmetry line if proper quadratic coefficients are used. An example is presented in Fig. \ref{fig2} (e)-(h) in which two TDPs appear (the coefficients can be found in the supplemental materials\cite{group_dis}). In this case, the Fermi level locates between the energies of the two TDPs. As shown in Fig. \ref{fig2} (f) and (g), the distributions of SBC in $k_z$ = 0 and $k_y$ = 0 planes are strong localization. However, positive SBC appear around TDP (Fig. \ref{fig2} (g)) and it is similar with the distribution of SBC around Weyl point\cite{TaAs}. The sign of SBC is opposite on the opposite sides of TDP, just like the characteristics of \emph{p} orbital. Furthermore, it is much like strong polarized \emph{p} orbital due to the large difference of band dispersions on the opposite sides of the TDP. The results also indicate that the negative contribution of SBC (blue zone in Fig. \ref{fig2} (g)) is greatly larger than that of positive one (red zone in Fig. \ref{fig2} (g)). Amazingly, the strong localized SBC also produce SHC platform as shown in Fig. \ref{fig2} (h) whose width is equal to the energy difference of the two group bands at A point.\\
%++++++++++++++++++++++++++++++++++++++++++++++++++++++++++++++++++++++++
\indent The conditions to produce SHC platform in metallic system can be briefly summarized as: (i) symmetry protection similar with that of triply degenerate semimetals; (ii) special states such as $|\frac{3}{2},\pm\frac{3}{2}\rangle$ and $|\frac{1}{2}, \pm\frac{1}{2}\rangle$\cite{group_dis} around the Fermi level; (iii) strong localization of SBC. Although such conditions can be easily fulfilled in a simplified $\textbf{k}\cdot\textbf{p}$ model, to realize the SHC platform in a real metallic material is still challenging. In present letter, we find that the experimentally synthesized $\theta$-TaN (we call it as TaN in following paper), a typical triply degenerate semimetal\cite{TaN2016,Nexus}, can achieve such SHC platform. We take TaN as a prototype to study the SHC using first-principles method\cite{group_dis}. The crystal structure of TaN \cite{TaN1972,TaN2010,TaN2012,TaN2014,TaN2016} is shown in Fig. \ref{fig1} (b) and (c). It is WC-type\cite{TaN2010} hexagonal crystal structure with space group $P\bar{6}m2 $ (No.187). The space group of TaN meets the requirement of the above condition (i).  The energy band structures of TaN with SOC are presented in Fig. \ref{fig3} (a). It is evident that the Fermi level crosses the bands around high symmetry K (K$^\prime$) and A points. There are two triply degenerate points on high symmetry line $\Gamma$-A below and above Fermi level, respectively. When the SOC is ignored, the states around Fermi level composed of one degenerate \emph{d$_{z^2}$} and double degenerate \emph{e$_g$} states (\emph{d$_{xy}$} and \emph{d$_{x^2-y^2}$}) of Ta, the results are in good agreement with previous report\cite{TaN2016}. The SOC effect will induce the \emph{d$_{z^2}$} and \emph{e$_g$} states to $|\frac{3}{2},\pm\frac{3}{2}\rangle$ and $|\frac{1}{2}, \pm\frac{1}{2}\rangle$\cite{group_dis} around the Fermi level, meeting the second requirement above. The band structures as shown in Fig. \ref{fig3} (a) indicate that the local extreme point along A-H for the two bands around Fermi level deviates from A point, which will greatly impact on the distribution of SBC around A point, the details will be discussed below. Fig. \ref{fig3} (b)-(d) show the different components of SHC with different concentrations of electron/hole doping. It is fascinating that platforms of SHC form for the four components $\sigma_{xy}^z$, $\sigma_{yx}^z$, $\sigma_{zy}^x$ and $\sigma_{zx}^y$, as shown in Fig. \ref{fig3} (b) and (d). The energy range of the platform reach up to 0.56 eV and the width is larger than quantized platform width in most previous report on 3D or 2D topological insulator. Such wide SHC platform is favorable for the system producing stable spin polarized current counter act with temperature fluctuation. The maximum energy of the platform is 0.03 eV lower than the intrinsic Fermi level, which indicates hole doping is necessary to constrain the Fermi level in the energy range of the platform. Considering the height of the platform, the $\sigma_{xy}^z$ and $\sigma_{yx}^z$ are more valuable for further application. Therefore, we will focus on the origin mechanism of $\sigma_{xy}^z$ platform below.\\
%+++++++++++++++++++++++++++++++++++++++++++++++++++++++++++++++++++++++++
\begin{figure}
% Requires \usepackage{graphicx}
\includegraphics[trim={0.0in 0.0in 0.0in 0.0in},clip,width=4.5in]{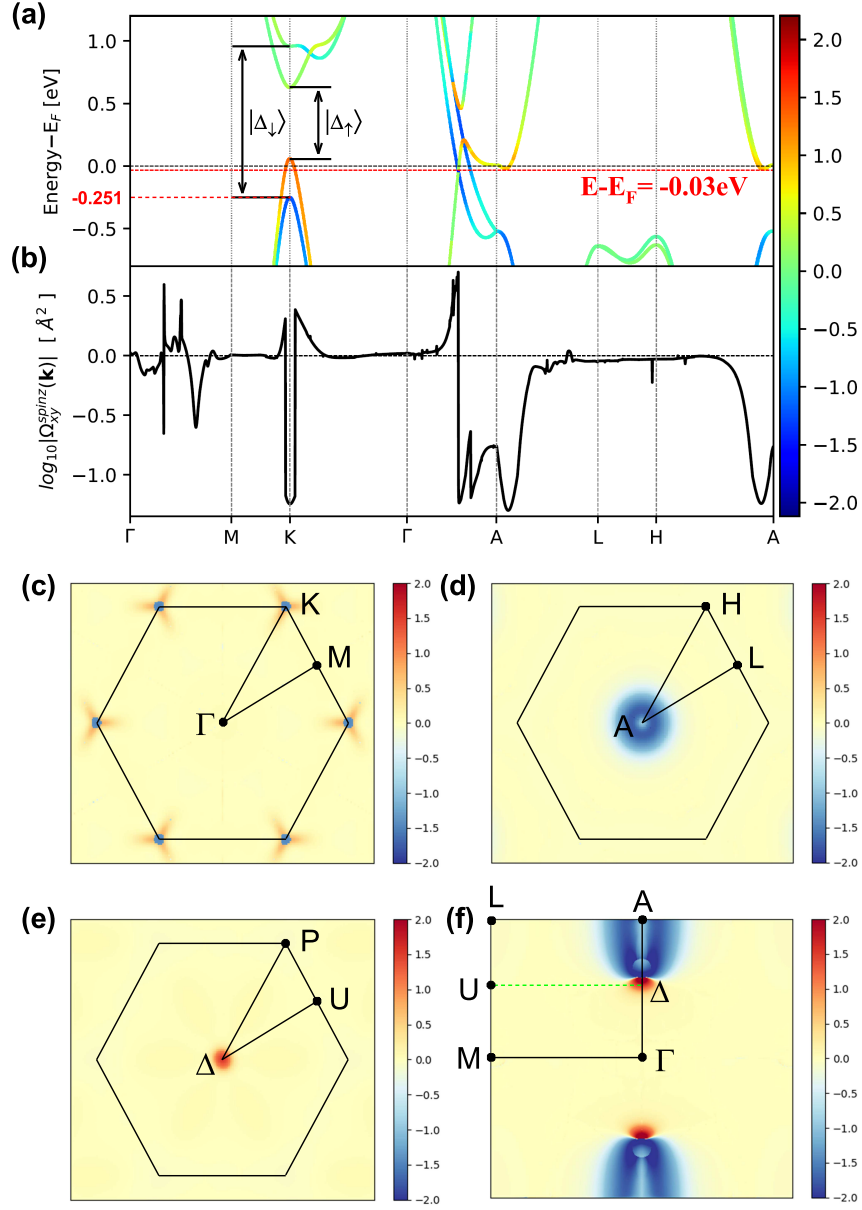}\\
\caption{(color online) (a) Band structures with the projection of SBC on a log scale. The red dashed horizontal lines represent E = E$_F$ - 0.03 eV. (b) The k-resolved SBC at E = E$_F$ - 0.03 eV. The k-resolved SBC on a log scale in a slice of the 2D BZ at k$_z$ = 0 (c), k$_z$ = $\pi$/c (d), k$_z$ = 0.52$\pi$/c (e), and k$_y$ = 0 (f)  for $\sigma_{xy}^z$ SHC of TaN at E = E$_F$ - 0.03 eV, respectively.}\label{fig4}
\end{figure}
%+++++++++++++++++++++++++++++++++++++++++++++++++++++++++++++++++++++++++
\indent As discussed above, the SHC platform should be related to the localization of the SBC of the system. We present the SBC along high symmetry lines and some sections of first Brilloune zone in Fig. \ref{fig4}. The results indicate that the SBC mainly concentrates on high symmetry line $\Gamma$-A,  A, and K points, especially around A point. Moreover, the SBC is isotropic in the plane perpendicular to the c axis around A point. As shown in Fig. \ref{fig4} (d), the extreme of SBC form a circular ring around A point, which is different from the results obtained from k${\cdot}$p model whose maximum of SBC just locate at A point. Such SBC distribution feature derives from the deviation of the band extreme around the Fermi level from A point, as shown in Fig. \ref{fig3} (a). The circular ring around A point can be well illustrated with formula (2) where the SBC inversely proportional to the square of the energy difference. As illustrated in Fig. \ref{fig3} (b), the SHC platform of $\sigma_{xy}^z$ is not absolutely flat in the energy range of [-0.59, -0.03] for TaN. The absolute height of the platform increases with the increasing of the Fermi level. The difference between absolute maximum (456 ($\hbar$/e) S/cm) and minimum (411 ($\hbar$/e) S/cm) of the SHC platform is 45 ($\hbar$/e) S/cm. The inclination of the SHC platform mainly comes from the complication of the band dispersion around TDPs. Along with the increasing of the Fermi level in the energy range of  [-0.59, -0.03], the SBC between A-L as shown in Fig. \ref{fig4} (b) is nearly unchanged. However, the average SBC on the high symmetry line of  $\Gamma$-A increases with the increase in the Fermi level due to the complicated energy band dispersion around the TDPs. As shown in Fig. \ref{fig4} (a), the color for two split $|\frac{3}{2}, \pm\frac{3}{2}\rangle$ bands on $\Gamma$-A are blue (negative). As the Fermi level move from -0.59 to -0.03 eV, more and more negative SBC contribute the total SHC. Especially at the energy of -0.03, two peaks of SBC can be observed on high symmetry line $\Gamma$-A. The evolution of the SBC around the TDPs directly produce the inclination of the SHC platform which is different from the absolute flat platform obtained from simplified $\textbf{k}\cdot\textbf{p}$ model. The results as shown in Fig. \ref{fig4} (b) and (c) indicate that besides around the TDPs there are large peaks of SBC around K and K$^\prime$. However the SBC peaks are severely restricted in a small region confined by the crossing line between the energy band around the K (K$^\prime$) and the Fermi level. The distribution feature can be found in Fig. \ref{fig4} (c). The contribution of the SBC around the K and K$^\prime$ points to the SHC is negligible. It is worth noting that the spin of the bands around K and K$^\prime$ point will split with each other due to the k-points are not time-reversal invariant. However, the time reversal and C$_3$ rotation symmetry will force the sign of SBC around K and K$^\prime$ to be the same\cite{mos2_2}. Generally, the point group of SBC is the subgroup of Berry curvature (BC)\cite{sbc_sym}. However, the time reversal symmetry correlation between SBCs around K and K$^\prime$ produces the original C$_3$ symmetry of BC into C$_6$ for SBC, as shown in Fig. \ref{fig4} (c). Moreover, the results as shown in Fig. \ref{fig4} (a) reveal that both valence and conduction bands are spin-splitting for TaN, which is different from the case of monolayer MoS$_2$\cite{mos2_1, mos2_2} that only valence bands are spin-splitting induced by spin-orbit coupling. Such spin-splitting of the low energy electronic states around K for TaN can be explained by a similar $\textbf{k}\cdot\textbf{p}$ Hamiltonian, and the results can be found in the supplemental materials\cite{group_dis}.\\
%++++++++++++++++++++++++++++++++++++++++++++++++++++++++++++
\indent Spin Hall angle (SHA) is another important parameter for SHE to characterize the conversion efficiency from the charge current to spin current. It is defined as the ratio of SHC to the charge conductivity, namely SHA=$ (e/\hbar)\sigma_{xy}^z /\sigma_{xx}$ where $\sigma_{xx}$ is the longitudinal charge conductivity, $\sigma_{xy}^z$ is the transverse SHC. Obviously, if we want to obtain a large SHA, large SHC as well as low charge conductivity are equally important. For pure 5\emph{d} metal Pt, although its SHC can reach 2400 $(\hbar/e)S/cm$, its SHA is small with the value of -0.07 because of large charge conductivity\cite{sha_1}. Considering the value of SHC platform of TaN is -456 $(\hbar/e)S/cm$ at the energy of -0.03 eV and the longitudinal charge conductivity is estimate to 736 $S/cm$\cite{group_dis}, the SHA of TaN reaches up to -0.62. The absolute SHA value of TaN is much larger than that of Pt and other reported values\cite{beta-W,MoTe2}, which indicates TaN is a high-performance material to produce pure spin current.\\
%++++++++++++++++++++++++++++++++++++++++++++++++++++++++++++
\indent In summary, with the four bands effective \textbf{k}${\cdot}$\textbf{p} model, we theoretically predict a SHC platform in the triply degenerate semimetals. The large SHC platform mainly come from the localized SBC produced between  $|\frac{3}{2},\pm\frac{3}{2}\rangle$ and
$|\frac{1}{2},\pm\frac{1}{2}\rangle$ states. It is worth to mention that the platform is robust against the TDP around Fermi level. With the example of TaN, we find, although the low energy electronic states of concrete materials maybe more complicated, the SHC platform can be well kept. The large SHC platform make TaN to be excellent material to produce stable spin current in a large of doping range. Moreover, we find that in the TaN material family, TaP, TaAs, TaBi, and TaSb show similar SHC platform, the results can be found in the supplemental materials\cite{group_dis}. The wide SHC platform of TaN material family is potential to spintronics devices with stable spin current under high temperature. \\
%============================================================
\begin{acknowledgments}
This work is supported by the National Natural Science Foundation of China (Grant No. 11804287, 11574260), Hunan Provincial Natural Science Foundation of China (2019JJ50577) and the Scientific Research Fund of Hunan Provincial Education Department (18A051).
\end{acknowledgments}
$\ddagger$ These authors contributed equally to this work.
%============================================================
%\bibliography{zou}
%\begin{references}{1}

%\bibliography{apssamp}% Produces the bibliography via BibTeX.

\end{document}